\documentstyle[12pt]{article}
\normalbaselines

\begin{document}
\begin{flushright}
{\raggedleft

hep-lat/0002022\\[3cm]}

\end{flushright}

\renewcommand{\thefootnote}{\fnsymbol{footnote}}
\begin{center}
{\LARGE\baselineskip0.9cm 
Isocliny in spinor space and Wilson fermions\\[1.5cm]}

{\large K. Scharnhorst\footnote[2]{E-mail:
{\tt scharnh@physik.hu-berlin.de}}
}\\[0.3cm]

{\small Humboldt-Universit\"at zu Berlin

Institut f\"ur Physik

Invalidenstr.\ 110

10115 Berlin

Federal Republic of Germany}\\[1.5cm]

\begin {abstract}
We show that Clifford algebras are closely
related to the study of isoclinic subspaces of spinor spaces
and, consequently, to the Hurwitz-Radon matrix problem.
Isocliny angles are introduced to parametrize gamma matrices, i.e.,
matrix representations of the generators of finite-dimensional 
Clifford algebras $C(m,n)$. Restricting the consideration to the
Clifford algebra $C(4,0)$, this parametrization is then applied 
to the study of Dirac traces occurring in Euclidean lattice quantum 
field theory within the hopping parameter expansion for Wilson fermions.
\end{abstract}

\end{center}

\renewcommand{\thefootnote}{\arabic{footnote}}

\thispagestyle{empty}

\newpage
\section{Introduction}

Clifford algebras \cite{port}-\cite{loun}
play an important role in a number of branches 
of mathematics and physics and are widely studied, therefore.
However, the initial motivation for the present study has arisen from
Euclidean lattice quantum field theory and, in particular, from 
the investigation of certain problems related to Wilson fermions
which will be described in greater detail further below.
Here, it suffices to mention that within the hopping parameter
expansion for Wilson fermions Dirac traces (in $d=2,3,4$ lattice dimensions 
related to the Clifford algebra $C(d,0)$) have to be calculated
whose detailed qualitative and quantitative understanding 
is of considerable theoretical importance. While in $d=2$ 
a rather complete understanding exists, it is fairly
incomplete for the physically more interesting dimensions $d=3,4$.
Enquiries into the existent mathematical and physical literature showed that 
researchers in the field of Clifford algebras had not developed
so far a framework which would have been sufficiently suited 
for the further exploration of this subject. In attempting to fill
this gap soon it turned out that the relevant structures are not
special to the Clifford algebras $C(d,0)$, $d=2,3,4$, but can be
identified, at least in some preliminary way, 
for any finite-dimensional Clifford algebra $C(m,n)$.
Therefore, the present study has two parts -- a mathematical one
which is focussed on Clifford algebras and another one which 
applies the obtained mathematical insight to a theoretical 
physics problem, i.e., the study of Wilson fermions.\\

In the first (mathematical)
part (sect.\ 2), we consider an arbitrary finite-di\-men\-sio\-nal 
Clifford algebra $C(m,n)$ and show in subsect.\ 2.2, 
after having given the necessary 
notations in subsect.\ 2.1, how it is related to isoclinic
subspaces of spinor space. This subject is considered here for the 
first time in the literature (some initial study of the subject 
by the present author for the Clifford algebra $C(3,0)$ can be found
in \cite{scha4}). While the applied method is completely satisfactory
for the Clifford algebra $C(4,0)$ which we are primarily interested
in from a theoretical physics point of view the investigation to be 
presented remains preliminary to some extent, as is explained in detail in 
subsect.\ 2.2, for the general case of an arbitrary finite-dimensional 
Clifford algebra $C(m,n)$. As one main result, we find that
the eigenspaces of the generators of the Clifford algebra
$C(m,n)$ to the eigenvalues $\lambda = \pm 1$, or $\lambda = \pm i$, 
are isoclinic to each other. Specifically,
eigenspaces belonging to different generators are
isoclinic to each other with an angle $\Theta=\pi/4$.
This result is then further exploited in subsect.\ 2.3
to set up a parametrization of gamma matrices, i.e., matrix 
representations of the generators of the Clifford algebra
$C(m,n)$, in terms of isocliny angles. According to Wong
\cite{wong1}, the description of isoclinic subspaces
is closely related to the Hurwitz-Radon matrix problem
and, consequently, we exploit this link in subsect.\ 2.4 
to connect Clifford algebra representations to the Hurwitz-Radon 
matrix problem. Finally, in subsect.\ 2.5 we discuss some 
formula which will turn out to be useful for the study 
of Wilson fermions.\\ 

The second (theoretical physics) part
(sect.\ 3) of the article is devoted to the study of
Wilson fermions in Euclidean lattice field theory.
Fermions play a major role in most physically relevant
models of quantum field theory, but their inclusion
into numerical studies within lattice field theory
remains to be hampered by the so-called sign problem, i.e.,
the appearance of contributions of alternating sign 
which have to be summed up. It is clear that in principle this
sign problem is related to the anticommuting and spinorial 
character of fermionic variables. But, concentrating our attention
onto Wilson fermions, not much is known in qualitative and 
quantitative respect concerning the emergence of sign factors
in theories containing them. As mentioned above, 
the situation is understood rather satisfactory in 2 lattice dimensions
\cite{stam}, but an equivalent understanding in 3 and 4 dimensions
is lacking. It is highly desirable for two reasons at least.
In recent years, to a large extent based on the insight 
obtained into the sign problem
for Wilson fermions in 2 dimensions exact equivalences between 
purely fermionic models of lattice quantum field theory and (multi-color) 
loop models (with a bending rigidity $\eta =1/\sqrt{2} =\cos\Theta$)
of standard statistical mechanics
which can equivalently be understood as vertex models have 
been established \cite{salm}-\cite{gatt4} (also note 
\cite{gatt5,gatt3}). They represent a novel link
between both branches of theoretical physics.
These equivalent models are free of any sign 
problem and are interesting from an analytical as well as from 
a numerical point of view. It would clearly be of great 
theoretical interest to extend this novel link between 
lattice quantum field theory and standard statistical mechanics
to models in more than 2 dimensions. Second, the recently 
proposed meron-cluster algorithm allows to effectively approach
the numerical simulation of various models which exhibit a sign problem
\cite{chan,cox}. However, it is based on a detailed understanding
of the emergence of the sign factors in any model under consideration. 
This clearly adds further motivation for studying the sign problem
for Wilson fermions. In the first subsection of sect.\ 3 
we review the present state of knowledge
concerning the calculation of Dirac traces for Wilson fermions.
Based on the results obtained in sect.\ 2, in subsect.\ 3.2 
we then show that in 3 lattice dimensions the calculation of Dirac 
traces can be performed in a very simple manner on the basis of 
different choices for the representation of gamma matrices.
Subsect.\ 3.3 finally discusses the possible use of these 
representations for the calculation of Dirac traces in 4 dimensions.
The article closes with some discussion in sect.\ 4.\\

\renewcommand{\theequation}{\mbox{\arabic{section}.\arabic{equation}}}
\setcounter{equation}{0}
\section{Isoclinic subspaces of spinor spaces and Clifford algebras}

\subsection{Definitions and basics}

The generators 
of a (finite-dimensional) Clifford algebra $C(m,n)$ obey 
the standard relation
\begin{eqnarray}
\label{CL1}
\gamma_\mu \gamma_\nu\ +\ \gamma_\nu \gamma_\mu &=& 
2\ g_{\mu\nu}\ {\bf 1}
\end{eqnarray}
where $g_{\mu\nu}$, $\mu,\nu = 1,..,(m+n)$ are the elements of 
the diagonal matrix $g$ with $g_{\mu\mu} = 1$ for $1\le\mu\le m$
and $g_{\mu\mu} = -1$ for $m<\mu\le n$ 
and ${\bf 1}$ is the unit element of the Clifford algebra.
They act as operators in a space $V$ which is called the spinor space.
For our purposes we consider the generators $\gamma_\mu$ as 
$2s\times 2s$, $s\in {\bf N}$, matrices (gamma matrices) with complex 
entries (for our purposes, it does not matter here if 
these matrices correspond to irreducible matrix
representations of the Clifford algebra $C(m,n)$, or not). 
We choose the gamma matrices to be hermitian or antihermitian,
respectively,
\begin{eqnarray}
\label{CL2}
\gamma_\mu &=& g_{\mu\mu} \gamma_\mu^\dagger\ .
\end{eqnarray}
This is always possible (see, e.g., \cite{zhel}, chap.\ 1, \S 1, pp.\ 20-22,
\cite{choq2}, chap.\ I, subsect.\ 4.3, iii), p.\ 11).
We also define the following $2s\times 2s$ diagonal matrices (${\bf 1}_s$
denotes the $s\times s$ unit matrix).
\begin{eqnarray}
\label{CL3a}
\gamma_{E_p}&=&
\left(
\begin{array}{*{2}{c}}
{\bf 1}_s&0\\ 
0&-{\bf 1}_s
\end{array}
\right)\ =\ \gamma_E
\ \ ,\ \ \ \ \gamma^2_{E_p}\ = \ \ {\bf 1}\\
\label{CL3b}
\gamma_{E_n}&=&i \gamma_{E}\hspace{3.35cm},
\ \ \ \ \gamma^2_{E_n}\ = -{\bf 1}
\end{eqnarray}
Consider now the following eigenvalue
equation for vectors $\phi$ in the (finite-dimensio\-nal) complex vector 
space $V = V_{\bf C}$  ($\simeq {\bf C}_{2s}$, $s\in {\bf N}$).
\begin{eqnarray}
\label{CL4}
\gamma_\mu \phi&=& \lambda_\mu \phi
\end{eqnarray}
From eq.\ (\ref{CL1}) immediately follows 
$\lambda_\mu = \lambda_{\mu\pm} = \pm 1$ for $1\le\mu\le m$ and 
$\lambda_\mu = \lambda_{\mu\pm} = \pm i$ for $m<\mu\le n$. The eigenspaces
to the eigenvalues $\lambda_\mu = \pm 1, \pm i$ are 
of equal dimensionality $s$ (multiplying eq.\ (\ref{CL4})
by $\gamma_\nu$, $\nu\not=\mu$ and taking into account
eq.\ (\ref{CL1}) one finds that for any eigenfunction $\phi$
to the eigenvalue $\lambda$, $\gamma_\nu\phi$ is an 
eigenfunction to the eigenvalue $-\lambda$). Any gamma matrix 
$\gamma_\mu$ can be diagonalized my means of an unitary transformation
$U_\mu$,
\begin{eqnarray}
\label{unita}
U_\mu \gamma_\mu U^\dagger_\mu&=&\gamma^\prime_\mu\ , 
\end{eqnarray}
and be brought either to the form of $\gamma_{E_p}$, (\ref{CL3a}) 
(for $1\le\mu\le m$), 
or $\gamma_{E_n}$, (\ref{CL3b}) (for $m<\mu\le n$).\\

We construct now orthogonal projectors $P_{\mu\pm}$,
\begin{eqnarray}
\label{CL5a}
P^2_{\mu\pm} &=& P_{\mu\pm}\ ,\\
\label{CL5b}
P_{\mu\pm} &=& P^\dagger_{\mu\pm}\ ,
\end{eqnarray}
onto the $s$-dimensional eigenspaces $V_{\mu\pm}$ of the generators 
$\gamma_\mu$ of the Clifford algebra $C(m,n)$.
Explicitly, these projectors read
\begin{eqnarray}
\label{CL6}
P_{\mu\pm}&=& 
\frac{1}{2}\ \left( {\bf 1} +\frac{\gamma_\mu}{\lambda_{\mu\pm}}\right)\ .
\end{eqnarray}
The property (\ref{CL5b}) is fulfilled by virtue of eq.\ (\ref{CL2}).
In the same way, we also define orthogonal projection operators related to the 
eigenspaces $E_\pm$ of the matrices (\ref{CL3a}), (\ref{CL3b}).
\begin{eqnarray}
\label{CL7a}
P_{E_p\pm}&=& \frac{1}{2}\ \left( {\bf 1} \pm \gamma_{E_p}\right)\ =\
P_{E\pm}\\
\label{CL7b}
P_{E_n\pm}&=&\frac{1}{2}\ \left( {\bf 1} \mp i \gamma_{E_n}\right)\ =\
P_{E\pm}
\end{eqnarray}

\subsection{Isocliny}

In the following, we will be interested in certain geometric, 
more precisely angular 
relations between the eigenspaces of the gamma matrices $\gamma_\mu$,
$\gamma_E$. For the purpose of the present
investigation, we equip the spinor space $V = V_{\bf C}$ with the conventional 
(in general, $Spin$ non-invariant) Hermitian form (product)
\begin{eqnarray}
\label{CL8}
(a,b)_{\bf C} &=& \sum_{k=1}^{2s} \bar{a}_k b_k\ ,\ \ \  a, b \in V
\end{eqnarray}
($\bar{a}_k$ denotes the complex conjugate of $a_k$).
Before we proceed further, in this context the following comment is due.
In general, the inner product in a spinor space is 
appropriately chosen as being invariant
under the $Spin$ group related to the Clifford algebra under
study (\cite{kast}, app.\ B, \S 4, p.\ 307,
\cite{loun2}-\cite{coqu}, \cite{benn}, chap.\ 2, sect.\ 2.6,
p.\ 62, \cite{budi}, \cite{loun}, sect.\ 18, p.\ 231). 
However, the mathematical (geometrical) formalism we want to rely on 
in the present paper has only be developed so far in the literature
for a scalar product which is related to the Hermitian form (\ref{CL8}).
Primarily, at the moment we are interested in using this formalism
for deriving some information useful for certain theoretical
physics problems. For those Clifford algebras we are 
interested in from a theoretical physics point of view
the Hermitian product (\ref{CL8}) is invariant under the corresponding
$Spin$ group (we are primarily interested in $C(4,0)$ which 
$Spin(4,0) = SU(2)\otimes SU(2)$ is related to). 
Therefore, without any further apology 
the following investigation is based on
the Hermitian product (\ref{CL8}). But, one has to keep in mind
that all considerations below will have to be reconsidered
on the basis of $Spin$ invariant scalar
products for the general case of a Clifford algebra $C(m,n)$, if 
one wants to treat the most general case on the appropriate footing
in the future.\\

As already mentioned above, we are interested in the geometry of the set
of $s$-dimensional subspaces given by the eigenspaces of $\gamma_\mu$,
$\gamma_E$ in the $2s$-dimensional spinor space
$V$. Their geometric relation can be characterized by means of the so-called
{\it stationary angles (principal angles)} between any two of these 
subspaces which we will 
study in the present paper (for a comprehensive list of references on the 
subject of angles between subspaces see sect.\ II of \cite{scha4}).
In general, the relative position of two $s$-dimensional 
subspaces of an affine space which have a least one point in common 
can be characterized by means of $s$ stationary angles $\theta_k$,
$0\le\theta_k\le\frac{\pi}{2}$, $k=1,\ldots,s$ (we consider the spinor 
space $V$ as a vector space attached to the point 0 of some $2s$-dimensional
affine space). There exist different approaches to calculate these stationary
angles. Here, we rely on a different approach than that used in \cite{scha4}
in studying the Clifford algebra $C(3,0)$. The present approach
turns out to be more convenient for the consideration of a general Clifford
algebra $C(m,n)$. Consider the $2s\times 2s$ matrix 
\begin{eqnarray}
\label{CL9}
&P_U P_W P_U&
\end{eqnarray}
where $P_U$, $P_W$ are orthogonal projectors onto the $s$-dimensional
subspaces $U$, $W$ of some $2s$-dimensional space $V$ (of course, 
equally well one can consider the matrix $P_W P_U P_W$). The spectrum
of the matrix (\ref{CL9}) is given by $2s$ numbers $s$ numbers of which are
equal to 0 and the remaining $s$ numbers are given by $\cos^2\theta_k$,
$k=1,\ldots,s$ (\cite{shir1}, chap.\ IV, \S 31, pp.\ 393/394,
\cite{eato}, chap.\ 1, problem 32, p.\ 68). If the $s$ stationary
angles $\theta_k$, $k=1,\ldots,s$, are equal to each other the two 
subspaces $U$ and $W$ are called {\it isoclinic} (to each other) 
and it holds (\cite{guns}, sect.\ 2, p.\ 299, eq.\ (2.3), 
\cite{lemm}, p.\ 99, 2.3(i), \cite{shap}, sect.\ 1, p.\ 481, (1.2)(5))
\begin{eqnarray}
\label{CL10}
P_U P_W P_U&=&\cos^2\theta\ P_U
\end{eqnarray}
where $\theta$ is the isocliny angle ($\theta = \theta_1 =\ldots
=\theta_s$). In other words, two subspaces $U, W\subset V$ 
(of equal dimension) are isoclinic to 
each other if the angle $\theta$ between any vector $x\in U$ and its
orthogonal projection onto $W$ (with respect to some inner product,
here: the conventional Hermitian product) 
is independent of the vector $x$ (for a comprehensive list
of references on isoclinic subspaces see \cite{scha4}, sect.\ II) 
Note that the definition of angles in complex vector spaces 
requires special attention. For a corresponding review 
including the concept of isocliny angles see \cite{scha5}.
For our purposes it suffices to mention that the isocliny angle $\theta$
is independent of whether it is calculated in the complex vector
space $V=V_{\bf C}$ itself or in the corresponding  
real vector space $V_{\bf R}$
equipped with an almost complex structure $J$ and isometric to $V_{\bf C}$.\\

Having introduced some mathematical concepts in the preceding 
paragraph \linebreak which will be required for the further 
discussion we can now proceed with 
our study of the Clifford algebra $C(m,n)$.
Taking into account eq.\ (\ref{CL1}), a simple calculation 
yields the following result in the case of an arbitrary Clifford
algebra $C(m,n)$ ($\mu\not=\nu$, $k,l = \pm (1)$; incidentally,
this result should be expected to apply also for infinite-dimensional 
Clifford algebras)\footnote{Note that
eq.\ (\ref{CL11}) is formally shape invariant with respect to 
transformations $\Lambda\in Spin(m,n)$: $\gamma_\mu\longrightarrow
\gamma^\prime_\mu = \Lambda\gamma_\mu\Lambda^{-1}$. However, as
in general $\Lambda^{-1}\not= \Lambda^\dagger$ the projectors
defined in eq.\ (\ref{CL6}) are in the general case no 
longer orthogonal projectors
because $\gamma^\prime_\mu$ is no longer hermitian or antihermitian
and, therefore, property (\ref{CL5b}) no longer applies.}.
\begin{eqnarray}
\label{CL11}
P_{\mu k} P_{\nu l} P_{\mu k}
&=&\tau\ P_{\mu k}\ ,\ \ \ \tau\ =\ \frac{1}{2}
\end{eqnarray}
By virtue of eq.\ (\ref{CL10}) this means that any two 
eigenspaces of two different gamma matrices $\gamma_\mu$,
$\mu = 1,\ldots,(m+n)$, are isoclinic to each other with
the isocliny angle
\begin{eqnarray}
\label{CL12}
\Theta &=&\frac{\pi}{4}\ ,\ \ \ \cos^2\Theta\ =\ \frac{1}{2}\ =\ \tau 
\end{eqnarray}
(the two different eigenspaces of any generator $\gamma_\mu$ are of 
course isoclinic to each other with an angle $\theta =\frac{\pi}{2}$).
This geometric relation has been considered in the literature for the 
first time in \cite{scha4} in the case of the real Clifford 
algebra $C(3,0)$.\\

As we want to rely in the following on the results obtained by Wong 
\cite{wong1}, until further notice we consider not the 
complex vector space $V$ itself (of (complex) dimension $2s$) but 
the real vector space $V_{\bf R}$ (of dimension
$4s$) associated with $V$ by means of an almost complex structure $J$
and isometric to $V$. To be specific, we define the 
almost complex structure $J$ by assigning any complex entry 
$z = a + i b$, $a,b\in {\bf R}$, of the gamma matrices $\gamma_\mu$ a 
$2\times 2$ matrix as follows.
\begin{eqnarray}
\label{CL13}
z\ =\ a + ib &\longrightarrow &
\left(
\begin{array}{*{2}{c}}
a&-b\\ 
b&a
\end{array}
\right)
\end{eqnarray}
Accordingly, until further notice we consider the gamma matrices $\gamma_\mu$
as $4s\times 4s$ matrices with real entries. 
Now, from eq.\ (\ref{CL11}) we may conclude that 
the set of eigenspaces of the gamma matrices $\gamma_\mu$,
$\mu = 1,\ldots,(m+n)$, forms a {\it set of mutually isoclinic
$2s$-planes} in ${\bf R}_{4s}$ (\cite{wong1}, part I, sect.\ 3, p.\ 19). 
The elements of such a set are pairwise isoclinic. Obviously, this set
is a subset of some {\it maximal set of mutually isoclinic 
$2s$-planes} $\Phi$.
Such a maximal set is defined by the property that it is not a proper
subset of a larger set of mutually isoclinic $2s$-planes.
Furthermore, a (as large as possible) set ($\subset\Phi$)
of subspaces of ${\bf R}_{4s}$ is called an {\it equiangular frame} if its
elements are pairwise isoclinic to each other with the angle $\pi/4$
(\cite{wong1}, part I, sect.\ 5, p.\ 40). Consequently, the 
$(m+n)$ eigenspaces of the gamma matrices $\gamma_\mu$ to the
eigenvalues $\lambda_{\mu +}$ or $\lambda_{\mu -}$, respectively, form
subsets of certain equiangular frames. 
For the purpose of the present paper it appears to be useful 
to consider two disjoint equiangular frames $\Omega$  ---
one ($\Omega_+$) related to the eigenspaces to the eigenvalue 
$\lambda_{\mu +}$, and the other one ($\Omega_-$) related to the eigenspaces
to the eigenvalue $\lambda_{\mu -}$. The following theorem by Wong 
will be helpful then ($\Phi$ is any maximal set of mutually isoclinic 
$2s$-planes in ${\bf R}_{4s}$; the indices have been changed to conform
to the notation used in the present article): ``If the angles 
between any $2s$-plane of $\Phi$ and the $p$ $2s$-planes of an 
equiangular frame are $\theta_k$ $(1 \le k \le p)$, then
\begin{eqnarray}
\label{CL14}
\sum^p_{k=1}\cos^2 2\theta_k&=&1\ \ .
\end{eqnarray}
Conversely, given any set of $p$ angles $\theta_k$ $(1 \le k \le p)$
such that $0 \le \theta_k \le \pi$ and $\sum \cos^2 2 \theta_k = 1$,
then there exists a unique $2s$-plane isoclinic to each of the $p$
$2s$-planes of a given equiangular frame, making angles $\theta_k$ with
them, and this $2s$-plane belongs to $\Phi$'' (\cite{wong1}, pt.\ I, 
sect.\ 5, p.\ 41, theorem 5.3 (b)).  We denote this isoclinic $2s$-plane
making angles $\theta_k$ $(1 \le k \le p)$ with the equiangular frame
$\Omega_+$ by $A_+$. It is always possible 
to give the related orthogonal projector $P_{A_+}$ the form
\begin{eqnarray}
\label{CL15}
P_{A_+}&=&P_{E_+}
\end{eqnarray}
(\cite{wong1}, pt.\ I, sect.\ 3, p.\ 19). Together with $A_+$ also
its orthogonal complement $A_-$ (${\bf R}_{4s} = A_+ \oplus A_-$)
is an element of $\Phi$. The $2s$-plane $A_-$ makes the angles 
\begin{eqnarray}
\label{hat}
\hat{\theta}_k &=& \frac{\pi}{2} -\theta_k 
\end{eqnarray}
with the elements of $\Omega_+$
(\cite{wong1}, pt.\ I, sect.\ 2, p.\ 16, lemma 2.2)
and the orthogonal projector onto it reads in accordance with
eq.\ (\ref{CL15})
\begin{eqnarray}
\label{CL16}
P_{A_-}&=&P_{E_-}\ .
\end{eqnarray}
From these considerations we immediately conclude that in view of 
eq.\ (\ref{CL10}) the following equations apply.
\begin{eqnarray}
\label{CL17a}
P_{E_+} P_{\mu +} P_{E_+}&=&\cos^2\theta_\mu\ P_{E_+}\\
\label{CL17b}
P_{E_-} P_{\mu +} P_{E_-}&=&\cos^2\hat{\theta}_\mu\ P_{E_-}\ =\
\sin^2\theta_\mu\ P_{E_-}
\end{eqnarray}
From eqs.\ (\ref{CL17a}), (\ref{CL17b}), by simple algebra 
the analogous equations which are expected to apply for symmetry 
reasons follow.
\begin{eqnarray}
\label{CL18a}
P_{E_-} P_{\mu -} P_{E_-}&=&\cos^2\theta_\mu\ P_{E_-}\\
\label{CL18b}
P_{E_+} P_{\mu -} P_{E_+}&=&\cos^2\hat{\theta}_\mu\ P_{E_+}\ =\
\sin^2\theta_\mu\ P_{E_+}
\end{eqnarray}
In other words, the isoclinic $2s$-plane $A_-$
makes angles $\hat{\theta}_k$ $(1 \le k \le p)$ 
with the elements of the equiangular frame $\Omega_-$.
For a two-dimensional illustration of the eqs.\ 
(\ref{CL17a})-(\ref{CL18b}) see fig.\ 1.\\

\begin{table}[t]
\unitlength1.mm
\begin{picture}(150,90)
\put(35,5){
\begin{picture}(80,80)
\linethickness{0.15mm}
\put(0,40){\line(1,0){80}}
\put(40,0){\line(0,1){80}}
\put(40,40){\line(5,2){40}}
\put(40,40){\line(-5,-2){40}}
\put(40,40){\line(2,-5){16}}
\put(40,40){\line(-2,5){16}}
\put(40,40){\oval(15,15)[tr]}
\put(44,50){\parbox[b]{10mm}{$\hat{\theta}_\mu$}}
\put(54,41.5){\parbox[b]{10mm}{$\theta_\mu$}}
\put(82,38){\parbox[b]{10mm}{$E_+$}}
\put(38.5,82){\parbox[b]{10mm}{$E_-$}}
\put(82,55){\parbox[b]{25mm}{$V_{\mu +}$}}
\put(22,82){\parbox[b]{25mm}{$V_{\mu -}$}}
\end{picture}  }
\end{picture}

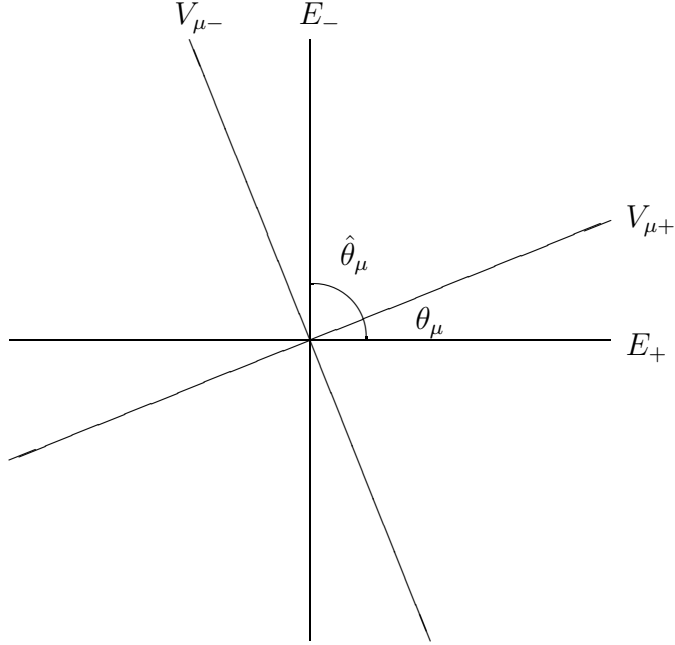
{\bf Figure 1:} Geometry of the eigenspaces $V_{\mu\pm}$ of the gamma
matrix $\gamma_\mu$ and of the subspaces $E_{\pm}$
\end{table}

\subsection{Parametrization of gamma matrices in terms of isocliny angles}

Having established the equations (\ref{CL17a})-(\ref{CL18b}) from now 
on we can return in our consideration to the original spinor space $V$
and derive from these equations further conclusions. To begin with, by
taking the sum of the left and right hand sides of 
the eqs.\ (\ref{CL17a}), (\ref{CL18a}) one can obtain the following 
relation by inserting eqs.\ (\ref{CL6})-(\ref{CL7b})
(the same result follows from eqs.\ (\ref{CL17b}), (\ref{CL18b});
in the following equations me omit the unit matrices ${\bf 1}$, 
${\bf 1}_s$ whenever it seems to be appropriate).
\begin{eqnarray}
\label{CL19}
\gamma_\mu \gamma_E\ +\ \gamma_E \gamma_\mu&=&2\lambda_{\mu +}
\cos 2\theta_\mu
\end{eqnarray}
Taking the difference of the left and right hand sides of 
the eqs.\ (\ref{CL17a}), (\ref{CL18a}) yields 
\begin{eqnarray}
\label{CL20a}
W_{\mu +} \gamma_\mu W^\dagger_{\mu +}&=&\lambda_{\mu +}
\cos^2\theta_\mu\ \gamma_E\ ,\\
\label{CL20b}
W_{\mu +}&=&\frac{1}{2}\ \left( {\bf 1} +
\frac{\gamma_E \gamma_\mu}{\lambda_{\mu +}}\right)\ .
\end{eqnarray}
Correspondingly,
taking the difference of the left and right hand sides of 
the eqs.\ (\ref{CL17b}), (\ref{CL18b}) yields 
\begin{eqnarray}
\label{CL21a}
W_{\mu -} \gamma_\mu W^\dagger_{\mu -}&=&\lambda_{\mu -}
\sin^2\theta_\mu\ \gamma_E\ =\ 
\lambda_{\mu -}\cos^2\hat{\theta}_\mu\ \gamma_E\ ,\\
\label{CL21b}
W_{\mu -}&=&\frac{1}{2}\ \left( {\bf 1} +
\frac{\gamma_E \gamma_\mu}{\lambda_{\mu -}}\right)\ .
\end{eqnarray}
As one sees, the equations related to $W_{\mu +}$ and $W_{\mu -}$
just differ in the factor $\lambda_{\mu\pm}$ and the use of the 
angles $\theta_\mu$ and $\hat{\theta}_\mu$. From the eqs.\ 
(\ref{CL20b}), (\ref{CL21b}) one finds
\begin{eqnarray}
\label{CL22}
W_{\mu -}\ +\ W_{\mu +}&=&{\bf 1}\ .
\end{eqnarray}
The eqs.\ (\ref{CL20b}), (\ref{CL21b}) entail
\begin{eqnarray}
\label{CL23a}
W_{\mu\pm} \gamma_E&=&\gamma_E W^\dagger_{\mu\pm}\ ,\\
\label{CL23b}
\gamma_\mu W_{\mu\pm}&=&W^\dagger_{\mu\pm} \gamma_\mu\ , 
\end{eqnarray}
and
\begin{eqnarray}
\label{CL24}
\gamma_\mu&=&\lambda_{\mu\pm}\ \gamma_E\ 
\left( 2 W_{\mu\pm} - {\bf 1}\right)\ .
\end{eqnarray}

%\label{CL25}
%\label{CL26}

We can now infer further properties of the matrices $W_{\mu\pm}$ 
from those of the gamma matrices $\gamma_\mu$.
Taking into account eqs.\ (\ref{CL19}), (\ref{CL20b}), (\ref{CL21b}) we find
($\theta_{\mu +} =\theta_\mu$, 
$\theta_{\mu -} =\hat{\theta}_\mu =\frac{\pi}{2} - \theta_\mu$)
\begin{eqnarray}
\label{CL27}
W_{\mu\pm} W^\dagger_{\mu\pm} &=&W^\dagger_{\mu\pm} W_{\mu\pm}\ =\
\left\{
\begin{array}{*{1}{c}}
\cos^2\theta_\mu\\ 
\sin^2\theta_\mu
\end{array}
\right\}\ =\  \cos^2 \theta_{\mu\pm}\ .
\end{eqnarray}
From $\gamma^2_\mu =g_{\mu\mu}$ (eq.\ (\ref{CL1})) and 
eqs.\ (\ref{CL24}), (\ref{CL27}) we obtain
\begin{eqnarray}
\label{CL28}
W_{\mu\pm}\ +\ W^\dagger_{\mu\pm} &=&
2\ \left\{
\begin{array}{*{1}{c}}
\cos^2\theta_\mu\\ 
\sin^2\theta_\mu
\end{array}
\right\}\ =\ 
2\ \cos^2 \theta_{\mu\pm}\ .
\end{eqnarray}
Also, from eq.\ (\ref{CL1}) we find after inserting eq.\ (\ref{CL24})
and using eq.\ (\ref{CL28})
\begin{eqnarray}
\label{CL29}
W^\dagger_{\mu\pm} W_{\nu\pm}\ +\ W^\dagger_{\nu\pm} W_{\mu\pm}&=&
W_{\mu\pm} W^\dagger_{\nu\pm}\ +\ W_{\nu\pm} W^\dagger_{\mu\pm}\nonumber\\ 
&=&\cos^2\theta_{\mu\pm}\ +\ \cos^2\theta_{\nu\pm}\ +\
\frac{\delta_{\mu\nu} -1}{2}\ .
\end{eqnarray}

For the further discussion, it turns out to be convenient to 
represent the matrices $W_{\mu\pm}$ as block matrices.
We write
\begin{eqnarray}
\label{CL30}
W_{\mu\pm}&=&
\left(
\begin{array}{*{2}{c}}
w_{\mu\pm 11}&w_{\mu\pm 12}\\ 
w_{\mu\pm 21}&w_{\mu\pm 22}
\end{array}
\right)
\end{eqnarray}
where the submatrices $w_{\mu\pm kl}$ are $s\times s$ matrices.
From eq.\ (\ref{CL23a}), we immediately find
\begin{eqnarray}
\label{CL31a}
w^\dagger_{\mu\pm 21}&=&-\ w_{\mu\pm 12}\ ,\\
\label{CL31b}
w^\dagger_{\mu\pm 11}&=&w_{\mu\pm 11}\ \ \ \ ,\ \ \ 
w^\dagger_{\mu\pm 22}\ =\ w_{\mu\pm 22}\ .
\end{eqnarray}
Eq.\ (\ref{CL28}) then entails 
\begin{eqnarray}
\label{CL32}
w_{\mu\pm 11}\ = \ w_{\mu\pm 22} &=&
\left\{
\begin{array}{*{1}{c}}
\cos^2\theta_\mu\\ 
\sin^2\theta_\mu
\end{array}
\right\}\ =\ 
\cos^2 \theta_{\mu\pm}\ .
\end{eqnarray}
Consequently, the $W_{\mu\pm}$ matrices assume the explicit form
(for convenience, we introduce the notation
$w_{\mu\pm 12} = w_{\mu\pm}/2$).
\begin{eqnarray}
\label{CL33a}
W_{\mu\pm}&=&\frac{1}{2}\ 
\left(
\begin{array}{*{2}{c}}
2 \cos^2\theta_{\mu\pm} &w_{\mu\pm}\\ 
-w^\dagger_{\mu\pm}&2 \cos^2\theta_{\mu\pm}
\end{array}
\right)\ .
\end{eqnarray}
By virtue of eq.\ (\ref{CL22}) holds
\begin{eqnarray}
\label{CL35}
w_{\mu +}&=&-\ w_{\mu -} 
\end{eqnarray}
and eq.\ (\ref{CL27}) yields
\begin{eqnarray}
\label{CL36}
w_{\mu\pm} w^\dagger_{\mu\pm}&=&w^\dagger_{\mu\pm} w_{\mu\pm}\ =\
\sin^2 2\theta_\mu\ =\
\sin^2 2\hat{\theta}_\mu\ .
\end{eqnarray}
The gamma matrices $\gamma_\mu$ read in accordance
with eq.\ (\ref{CL24}) 
\begin{eqnarray}
\label{CL33b}
\gamma_\mu&=&\lambda_{\mu\pm}\ 
\left(
\begin{array}{*{2}{c}}
\cos 2\theta_{\mu\pm} &w_{\mu\pm}\\ 
w^\dagger_{\mu\pm}&-\cos 2\theta_{\mu\pm}
\end{array}
\right)\ .
\end{eqnarray}
This is the parametrization of gamma matrices in terms of 
isocliny angles mentioned in the Introduction. By virtue 
of eqs.\ (\ref{hat}), (\ref{CL35}), this 
representation is independent of whether on the r.h.s.\ of eq.\ 
(\ref{CL33b}) the upper or lower signs are chosen.\\

For future purposes, we express the following product of $W_{\mu\pm}$
matrices in terms of the matrices $w_{\mu\pm}$ ($k,l = \pm (1)$).
\begin{eqnarray}
\label{CL34a}
T_{(\mu k,\nu l)}&=& W_{\mu k} W^\dagger_{\nu l}\ =\
\left(
\begin{array}{*{2}{c}}
t_{(\mu k,\nu l) 11}&t_{(\mu k,\nu l) 12}\\ 
t_{(\mu k,\nu l) 21}&t_{(\mu k,\nu l) 22}
\end{array}
\right)\\
\label{CL34b}
t_{(\mu k,\nu l) 11}&=&\cos^2\theta_{\mu k} \cos^2\theta_{\nu l}\ +\
\frac{1}{4}\ w_{\mu k} w^\dagger_{\nu l}\\
\label{CL34c}
t_{(\mu k,\nu l) 22}&=&\cos^2\theta_{\mu k} \cos^2\theta_{\nu l}\ +\
\frac{1}{4}\ w^\dagger_{\mu k} w_{\nu l}\\
\label{CL34d}
t_{(\mu k,\nu l) 12}&=&-\ t^\dagger_{(\mu k,\nu l) 21}\ =\ \frac{1}{2}\ 
\left(\cos^2\theta_{\nu l}\ w_{\mu k}\ -\ 
\cos^2\theta_{\mu k}\ w_{\nu l}
\right)
\end{eqnarray}
Inserting eqs.\ (\ref{CL34a})-(\ref{CL34d}) into (\ref{CL29})
and taking into account eq.\ (\ref{CL35}) we finally obtain
\begin{eqnarray}
\label{CL37}
w^\dagger_{\mu k} w_{\nu l}\ +\ w^\dagger_{\nu l} w_{\mu k}&=&
w_{\mu k} w^\dagger_{\nu l}\ +\ w_{\nu l} w^\dagger_{\mu k}\nonumber\\ 
&=&2 
\left[ k\ l\ \delta_{\mu\nu} - 
\cos 2\theta_{\mu k} \cos 2\theta_{\nu l}\right]\ .
\end{eqnarray}
Introducing the matrix $\widetilde{w}_{\mu\pm}$ by writing
\begin{eqnarray}
\label{CL38}
w_{\mu\pm}&=&2\ \cos^2 \theta_{\mu\pm}\ \widetilde{w}_{\mu\pm}
\end{eqnarray}
one finds that the eqs.\ (\ref{CL36}) and (\ref{CL37}) expressed
in terms of $\widetilde{w}_{\mu\pm}$ exactly agree with the eqs.\ (3.1)
and (3.2) in \cite{wong1}, part I, sect.\ 3, p.\ 20, lemma 3.1
(in the notation of Wong the matrices $\widetilde{w}_{\mu\pm}$ 
describe isoclinic subspaces with an isocliny angle 
$\theta_{\mu\pm}$ relative to the subspace $E_+$).\\

\subsection{Clifford algebras and the Hurwitz-Radon matrix problem}

From eqs.\ (\ref{CL20a}), (\ref{CL21a}) we recognize that any gamma
matrix $\gamma_\mu$ can be diagonalized in accordance with eq.\ (\ref{unita}) 
my means of the unitary matrix (we disregard here the case 
$\theta_{\mu\pm} = \frac{\pi}{2}$ which needs to be discussed 
separately) 
\begin{eqnarray}
\label{CL39}
U_{\mu\pm}&=&\frac{W_{\mu\pm}}{\cos\theta_{\mu\pm}}\ .
\end{eqnarray}
Using eqs.\ (\ref{CL20b}), (\ref{CL21b}) and (\ref{CL2}) one finds 
the following representations of the gamma matrices in terms of the 
matrices $U_{\mu\pm}$.
\begin{eqnarray}
\label{CL40}
\gamma_\mu&=&\lambda_{\mu\pm} \left( U^\dagger_{\mu\pm}\right)^2 
\gamma_E\ =\ 
\lambda_{\mu\pm} U^\dagger_{\mu\pm} \gamma_E U_{\mu\pm}\ =\
\lambda_{\mu\pm} \gamma_E U^2_{\mu\pm}
\end{eqnarray}
Now, choose $\mu =\mu_0$ and elevate $U_{\mu_0 +}$ to a unitary matrix
by means of which all gamma matrices related to the Clifford algebra $C(m,n)$
are being transformed. Then, after some calculation taking 
into account eqs.\ (\ref{CL20b}), (\ref{CL21b}) and (\ref{CL19}) one finds 
(we apply here the inverse transformation compared with eq.\ (\ref{unita})) 
\begin{eqnarray}
\label{CL41}
\gamma_\nu&=&U^\dagger_{\mu_0 +} \gamma^\prime_\nu U_{\mu_0 +}\nonumber\\
&=&\gamma^\prime_\nu\ + \lambda_{\nu +}
\frac{\cos 2\theta_\nu}{\cos\theta_{\mu_0}}\ \gamma_E U_{\mu_0 +}
\ ,\ \ \ \nu\not=\mu_0\ .
\end{eqnarray}
As $\gamma^\prime_{\mu_0} =\lambda_{\mu_0 +} \gamma_E$ 
($\theta^\prime_{\mu_0} = 0$),
from eq.\ (\ref{CL14}) we immediately conclude that
\begin{eqnarray}
\label{CL42}
\theta^\prime_\nu&=&\Theta\ =\ \frac{\pi}{4}\ ,\ \ \ \nu\not=\mu_0\ .
\end{eqnarray}
Eq.\ (\ref{CL41}) then yields the following relation for the 
submatrices $w_{\nu +}$.
\begin{eqnarray}
\label{CL43}
w_{\nu +}&=&w^\prime_{\nu +}\ +\ 
\frac{\cos 2\theta_\nu}{2\cos^2\theta_{\mu_0}} 
\ w_{\mu_0 +}\ ,\ \ \ \nu\not=\mu_0
\end{eqnarray}
Inserting this relation into eq.\ (\ref{CL36}) we find
\begin{eqnarray}
\label{CL44}
w^{\prime\dagger}_{\nu +}\ w_{\mu_0 +}\ +\ 
w^\dagger_{\mu_0 +}\ w^\prime_{\mu +}&=&
w^\prime_{\nu +}\ w^\dagger_{\mu_0 +}\ +\ 
w_{\mu_0 +}\ w^{\prime\dagger}_{\nu +}\nonumber\\
&=& -\ 2 \cos 2\theta_\nu
\ ,\ \ \ \nu\not=\mu_0\ .
\end{eqnarray}
Consequently, 
once a set of gamma matrices $\gamma^\prime_\mu$, $\mu=1,\ldots,(m+n)$,
with $\theta^\prime_{\mu_0} = 0$ and $\theta^\prime_{\nu} = \frac{\pi}{4}$
for $\nu\not=\mu_0$ is given the choice of the matrix $w_{\mu_0 +}$
uniquely determines the set of gamma matrices $\gamma_\mu$ which 
exhibit the isocliny angles $\theta_\mu$ (and which are to be determined
from the eq.\ (\ref{CL44}) and for $\mu=\mu_0$ from eq.\ (\ref{CL36})).
According to eq.\ (\ref{CL37}), the $(m+n-1)$ matrices $w^\prime_{\mu +}$,
$\mu\not=\mu_0$ (of course, $w^\prime_{\mu_0 +}=0$), obey the equation
\begin{eqnarray}
\label{CL45}
w^{\prime\dagger}_{\mu +}\ w^\prime_{\nu +}\ +\ 
w^{\prime\dagger}_{\nu +}\ w^\prime_{\mu +}&=&
w^\prime_{\mu +}\ w^{\prime\dagger}_{\nu +}\ +\ 
w^\prime_{\nu +}\ w^{\prime\dagger}_{\mu +}\nonumber\\ 
&=&2 \delta_{\mu\nu}\ ,\ \ \ \mu,\nu\not=\mu_0 \ .
\end{eqnarray}
Eq.\ (\ref{CL45}) together with 
$w^\prime_{\mu +} w^{\prime\dagger}_{\mu +} = {\bf 1}_s$
(cf.\ eq.\ (\ref{CL36})) represent the {\it unitary Hurwitz-Radon matrix
problem} for the matrices $w^\prime_{\mu +} $ \cite{eckm1}, subsection 
1.4, p.\ 25, also see \cite{wong1}, part II, p.\ 67 and \cite{wong4}.
For a different approach relating (real) Clifford algebras to
(generalized) Hurwitz-Radon matrix problems see \cite{lawr1}-\cite{cnop2}.\\

\subsection{Some useful formula}

In this final subsection we focus our attention onto the matrix 
product (\ref{CL34a}). Let us start with the observation that
the matrices $t_{(\mu k,\nu l) ij}$, $i,j= 1,2$, obey the equation
\begin{eqnarray}
\label{CL46}
t_{(\mu k,\nu l) ij}\ t^\dagger_{(\mu k,\nu l) ij}&=&
t^\dagger_{(\mu k,\nu l) ij}\ t_{(\mu k,\nu l) ij}\nonumber\\
&=&\frac{1}{2}\ \left[1+ k\ l\ (-1)^{i+j}\delta_{\mu\nu}\right]\ 
\cos^2\theta_{\mu k}\ \cos^2\theta_{\nu l}\nonumber\\
&=&\tau\ \left[1+ k\ l\ (-1)^{i+j}\delta_{\mu\nu}\right]\ 
\cos^2\theta_{\mu k}\ \cos^2\theta_{\nu l}\ .
\end{eqnarray}
This can easily be checked by taking into account eq.\ (\ref{CL37}).
The last line can be seen to apply by writing 
\begin{eqnarray}
\label{CL47}
P_{\mu\pm} &=& U^\dagger_{\mu\pm} P_{E_+} U_{\mu\pm}
\end{eqnarray}
and inserting it
into eq.\ (\ref{CL11}). Now, for later purposes 
let us further study the matrix
$t_{(\mu k,\nu l) 11}$ for $\mu\not=\nu$. According to eq.\ 
(\ref{CL46}) the matrix 
\begin{eqnarray}
\label{CL48}
\check{t}_{(\mu k,\nu l)}
&=&\frac{t_{(\mu k,\nu l) 11}}{\sqrt{\tau}\ 
\cos\theta_{\mu k}\ \cos\theta_{\nu l}}\ 
\nonumber\\
&=&\sqrt{2}\ \cos\theta_{\mu k} \cos\theta_{\nu l}\ +\
\frac{w_{\mu k} w^\dagger_{\nu l}}{2 \sqrt{2} 
\cos\theta_{\mu k}\ \cos\theta_{\nu l}}
\end{eqnarray}
is unitary. Using eq.\ (\ref{CL37}), one can easily check that
this matrix can be represented  
as follows ($\alpha_{(\mu k,\nu l)}\in {\bf R}$). For
convenience, we introduce here an antisymmetric
sign factor $f_{\mu\nu} = - f_{\nu\mu} =\pm 1$ 
whose sign can later be arranged arbitrarily
to simplify explicit expressions.
\begin{eqnarray}
\label{CL49a}
\check{t}_{(\mu k,\nu l)}
&=&{\rm e}^{\displaystyle
\left[\alpha_{(\mu k,\nu l)}\ I_{(\mu k,\nu l)}\right]}\nonumber\\
&=&\cos\alpha_{(\mu k,\nu l)}\ +\ 
I_{(\mu k,\nu l)}\ \sin\alpha_{(\mu k,\nu l)}
\ ,\ \ \ \mu\not=\nu\\
\label{CL49b}
I_{(\mu k,\nu l)}&=&k\ l\ f_{\mu\nu}\ \sqrt{2}\ \ 
\frac{\left[w_{\mu k} w^\dagger_{\nu l} + 
\cos 2\theta_{\mu k} \cos 2\theta_{\nu l}\right]}
{\sqrt{-\cos 4\theta_{\mu k} - \cos 4\theta_{\nu l} }}
\ ,\nonumber\\
&&\ \ \ I^2_{(\mu k,\nu l)}\ =\ -\ {\bf 1}_s\\
\label{CL49c}
\sin\alpha_{(\mu k,\nu l)}&=&k\ l\ f_{\mu\nu}\ 
\frac{\sqrt{-\cos 4\theta_{\mu k} - \cos 4\theta_{\nu l} }}
{4\ \cos\theta_{\mu k}\ \cos\theta_{\nu l}}\\
\label{CL49d}
\cos\alpha_{(\mu k,\nu l)}&=&
\frac{1 + \cos 2\theta_{\mu k} + \cos 2\theta_{\nu l} }
{2 \sqrt{2}\ \cos\theta_{\mu k}\ \cos\theta_{\nu l}}
\end{eqnarray}
It seems to be worth emphasizing that the angle 
$\alpha_{(\mu k,\nu l)}$ only depends 
on the isocliny angles $\theta_{\mu k}$ and not on any
details of the choice of the matrices $w_{\mu k}$.
For $w_{\mu k} = w^\prime_{\mu k}$, i.e., 
$\theta_{\mu k} =\Theta\ =\ \frac{\pi}{4}$, 
$\mu\not=\mu_0$, from the above equations one immediately finds
\begin{eqnarray}
\label{CL50}
\check{t}_{(\mu k,\nu l)}&=&{\rm e}^{\displaystyle
\ \left[\frac{\pi}{4}\ w^\prime_{\mu k} w^{\prime\dagger}_{\nu l}
\right]}\ ,\ \ \ \mu\not=\nu,\ \mu, \nu\not=\mu_0\ .
\end{eqnarray}
Finally, note that the case $\theta_{\mu} = \frac{\pi}{2}$ requires
some special, separate consideration.\\

\setcounter{equation}{0}
\section{Dirac traces for Wilson fermions}

\subsection{The problem}

For a number of reasons, Wilson fermions are frequently 
applied in lattice field theory calculations 
(see, e.g., \cite{mont1,roth}).
The sign problem for fermions already shows up for
free fermions and for simplicity here we restrict our 
consideration to these.
The partition function $Z_\Lambda$ for free Wilson fermions 
on a $d$-dimensional (hyper-) cubic lattice $\Lambda$ is 
given by 
\begin{eqnarray}
\label{B1}
Z_\Lambda &=& \int D\psi D\bar\psi\ \ {\rm e}^{-S}\ \ \ ,
\end{eqnarray}
where $D\psi D\bar\psi$
denotes the multiple Grassmann integration on the lattice.
The action $S$ is defined by
\begin{eqnarray}
\label{B2}
S &=& \sum_{x\in \Lambda} \left( 
\sum_\mu \left(\bar\psi(x+ e_\mu)P_{\mu +} \psi(x) +
\bar\psi(x)P_{\mu -} \psi(x+ e_\mu)
\right) \right.\nonumber\\[0.3cm]
&&\ \ \ -\ M \bar\psi(x)\psi(x)\Bigg)
\end{eqnarray}
In eq.\ (\ref{B2}) the Wilson parameter $r$ has been set to 1. 
We are considering here Euclidean lattice field theory, therefore,
the Clifford algebra relevant for the study of Wilson fermions
is $C(d,0)$ ($d= 2,3,4$). Within the hopping parameter expansion the 
partition function $Z_\Lambda$ can be written as follows \cite{roth}, 
chap.\ 12, p.\ 165.
\begin{eqnarray}
\label{B3}
Z_\Lambda &=&M^{\vert\Lambda\vert}\ \exp\left[ -\sum_L\
(2\kappa)^{\vert L\vert}\ {\rm tr}\, \Gamma_L \right]
\end{eqnarray}
Here, $\kappa = 1/2M$ is the hopping parameter, $L$ is any single 
closed loop configuration on the lattice $\Lambda$, $\vert L\vert$ 
denotes its length and 
\begin{eqnarray}
\label{B4}
\Gamma_L &=&\prod_{l=1}^{\vert L\vert} P_{\mu_l k_l}\ =\
P_{\mu_1 k_1} P_{\mu_2 k_2} \cdots 
P_{\mu_{\vert L\vert -1} k_{\vert L\vert -1}} 
P_{\mu_{\vert L\vert} k_{\vert L\vert}}
\end{eqnarray}
is the path-ordered product of the projection matrices displayed in 
eq.\ (\ref{B2}) (in the hopping parameter expansion to each projector
$P_{\mu\pm}$ corresponds a (directed) line between two neighboring
points in the $d$-dimensional (hyper-) cubic lattice 
in the direction $\mu$). In the following we will just be interested in the 
calculation of the trace of an arbitrary such matrix $\Gamma_L$.
This problem has systematically been studied in \cite{stam}.\\

For 2 lattice dimensions, 
the problem has been solved in a completely satisfactory way
where one finds using $2s\times 2s$ gamma matrices, $s=1,2$, 
and applying free boundary conditions that
(\cite{stam}, p.\ 1131, eq.\ (15); also see \cite{scha3}, p.\ 489, eq.\ (50);
$B_L$ is the analogue of the Kac-Ward sign factor met in the 
study of the two-dimensional Ising model \cite{kac}, \cite{feyn},
chap.\ 5, sect.\ 5.4, p.\ 136)
\begin{eqnarray}
\label{B5}
{\rm tr}\, \Gamma_L &=&s\ B_L\ 2^{-C(L)/2}\ ,\ \ \ 
B_L\ =\ (-1)^{q(L)+1}\ .
\end{eqnarray}
Here, $C(L)$ is the number of corners 
and $q(L)$ is the number of self-intersections of the loop $L$.
Closely related to this result, free Wilson fermions described 
by the eqs.\ (\ref{B1}), (\ref{B2}) have been shown to be exactly 
equivalent in 2 dimensions to a two-color loop model with a bending rigidity
$\eta = 1/\sqrt{2}$ (\cite{scha3}, also see \cite{gatt2,gatt4}). 
For the partition function (\ref{B1}) holds in this case
\begin{eqnarray}
\label{B6}
Z_\Lambda&=&\left( Z_\Lambda\left[M,{1\over\sqrt{2}}\right]\right)^2\ \ ,
\end{eqnarray}
where $Z_\Lambda[z,\eta]$ is the partition function for
the self-avoiding loop model with mo\-no\-mer weight $z$
and a bending rigidity $\eta$ as defined in \cite{scha1}.
Within the hopping parameter expansion of the partition function
$Z_\Lambda$  (eq.\ (\ref{B1})) the equations (\ref{CL11}), (\ref{CL12}) can be 
represented pictorially as shown in fig.\ 2
(in \cite{gatt3}, sect.\ 4, fig.\ 2, p.\ 4860 
the relation (\ref{CL11}) has been dubbed ``kink rule'', 
also note \cite{gatt1}, sect.\ 2.2, fig.\ 3, p.\ 4558
and eq.\ (2.16), p.\ 4557). 
\begin{figure}[t]
\unitlength1.mm
\begin{picture}(150,10)
\linethickness{0.04cm}
\put(40,7){
\put(0,10){\line(1,0){10.2}}
\put(10,10){\line(0,-1){10}}
\put(9.8,0){\line(1,0){10.2}}
}
\put(100,7){
\put(0,5){\line(1,0){10}}
}
\put(5,21){
\parbox[t]{15cm}{
\begin{eqnarray}
\label{figtext1}
&&\hspace{-2mm}\hat{=}\hspace{1cm} \frac{1}{2}
\nonumber
\end{eqnarray}  }}
\end{picture}
{\bf Figure 2:} ``Kink rule'' to be used in calculating traces of
products of projection operators $P_{\mu\pm}$  which occur in 
the hopping parameter expansion of the partition function
$Z_\Lambda$, eq.\ (\protect\ref{B3}) (in this figure arrows 
have been omitted for simplicity). $\eta$ is the weight assigned to a 
corner in the loop model picture. 
\end{figure}
From this picture one immediately recognizes
that $\eta = \cos\Theta = 1/\sqrt{2}$ holds.\\

For 3 and 4 lattice dimensions, in the appendix of \cite{stam} 
the following result has been obtained. Apply a $4\times 4$
gamma matrix representation ($s=2$) of the Clifford algebra $C(5,0)$ with 
\begin{eqnarray}
\label{B7a}
\theta_\mu &=&\frac{\pi}{4}\ ,\ \ \ \mu = 1,2,3,4,\\
\label{B7b}
\theta_5&=&0\ ,\\
\label{B7c}
w_{k +}&=&w^\prime_{k +}\ =\  - i \sigma_k\ ,\ \ \  k\ =\ 1,2,3,\\
\label{B7d}
w_{4 +}&=&w^\prime_{4 +}\ =\ {\bf 1}_2
\end{eqnarray}
(this is the so-called chiral representation,
see, e.g., \cite{mont1}, appendix 8.1, p.\ 435; $\sigma_k$ are the 
standard Pauli matrices). Then, one finds 
\begin{eqnarray}
\label{B8a}
{\rm tr}\, \Gamma_L &=&2\ B_L\ 2^{-C(L)/2}\ ,\\
\label{B8b}
B_L&=&\frac{1}{2}\ {\rm tr}\, \prod^{C(L)}_{l=1}\ 
{\rm e}^{\displaystyle
\ \left[\frac{\pi}{4}\ w^\prime_{\mu_l k_l} 
w^{\prime\dagger}_{\mu_{l+1} l_{l+1}}
\right]}
\end{eqnarray}
($\mu_{C(L) + 1} = \mu_1$, $l_{C(L) + 1} = l_1$; of course, the number
of loop sides is equal to the number of corners, we associate the side
$\# l$ with the corner $\# l$). This result
can easily be related to the discussion performed in sect.\ 2.
Inserting eq.\ (\ref{CL47}) into eq.\ (\ref{B4})  and 
taking into account the eqs.\ (\ref{CL39}), (\ref{CL34a}), 
(\ref{CL48}) results in
\begin{eqnarray}
\label{B9}
{\rm tr}\, \Gamma_L &=&\tau^{C(L)/2}\ 
{\rm tr}\, \prod^{C(L)}_{l=1}\ \check{t}_{(\mu_l k_l,\mu_{l+1} l_{l+1})}\ .
\end{eqnarray}
With the choice (\ref{B7a})-(\ref{B7d}), one immediately 
recovers from eq.\ (\ref{B9}) the
eqs.\ (\ref{B8a}), (\ref{B8b}). Clearly, the eq.\ (A.3) in
the appendix of \cite{stam} can be recognized in the eqs.\ 
(\ref{CL39}) and (\ref{CL47}). Further, in \cite{stam} it has
been argued that the matrices 
\begin{eqnarray}
\label{B10} 
{\rm e}^{\displaystyle
\ \left[\frac{\pi}{4}\ w^\prime_{\mu_l k_l} 
w^{\prime\dagger}_{\mu_{l+1} l_{l+1}}
\right]}
\end{eqnarray}
belong to a $j=\frac{1}{2}$ representation of the double octahedral group 
$^2O_3$ (for some related review see, e.g., \cite{john}) . 
The trace in eq.\ (\ref{B8b}) is a character of the double 
octahedral group $^2O_3$ and, therefore, one finds that 
\begin{eqnarray}
\label{B11}
B_L&=&\cos\frac{\theta_L}{2}\ .
\end{eqnarray}
$\theta_L$ can either be a multiple of $\pi$ or $\frac{2\pi}{3}$ if 
$C(L)$ is even, or an odd multiple of $\frac{\pi}{2}$ if $C(L)$ is odd.
This can easily be understood by taking recourse to the isomorphy of the 
octahedral group $O_3$ to the symmetric group $S_4$. An even permutation
$\in S_4$ corresponds to a rotation of the three-dimensional cube by an 
angle of $\pi$ or $\frac{2\pi}{3}$ (classes $3C_2$, $8C_3$, 
$6C^\prime_2$ of $O_3$) while an odd permutation corresponds to 
a rotation by an angle of $\frac{\pi}{2}$ (class $6C_4$ of $O_3$).
Consequently, the matrix (\ref{B10}) corresponds to an odd 
permutation. This fact immediately provides us with 
the explanation for the mentioned
rule (for a more technical argument yielding the same result 
see the original paper \cite{stam}). Furthermore, for loops in 3 
lattice dimensions $\theta_L$ cannot be a multiple of $\frac{2\pi}{3}$ 
which is not at the same time a multiple of $\pi$.\\

While eqs.\ (\ref{B8a}), (\ref{B8b}), (\ref{B11}) 
provide us with interesting
information, for 3 and 4 lattice dimensions a qualitative insight
comparable with eq.\ (\ref{B5}) and generalizing it to these
dimensions is lacking so far (some suggestion made for 3 
dimensions in \cite{stam}, p.\ 1131, can be seen to be 
incorrect by studying some simple examples). 
But, as explained in the Introduction
it is highly desirable to solve this problem in some satisfactory 
way or, at least, to make further progress in this direction. 
One obstacle to gain further insight seems to be that the 
calculation of $B_L$ according to the eqs.\ 
(\ref{B8b}), (\ref{B11})
is related to some non-abelian group (i.e., to the double 
octahedral group $^2O_3$). Consequently, based on the results 
obtained in sect.\ 2 in the following we study the question 
if it is possible
to relate the calculation of $B_L$ to some abelian group
(or to a group -- take this with a grain of salt --  exhibiting
as little as possible non-abelian structure, or a as simple as 
possible such structure). For 3 lattice dimensions, we
will show in the next subsection that such an abelianization is indeed
possible. For 4 dimensions, the situation is considerably more involved
and will separately be discussed in subsect.\ 3.3.\\

\subsection{Dirac traces in 3 lattice dimensions}

In 3 lattice dimensions, we can either choose $2\times 2$ ($s=1$)
or $4\times 4$ ($s=2$) gamma matrices to represent the Clifford 
algebra $C(3,0)$. We will start with considering 
$s=1$ and derive from this case all necessary information for
$s=2$. To clarify the relation between the Dirac traces taken in
the $s=1$ and $s=2$ representations of the Clifford algebra $C(3,0)$
let us start with the following observations (for the moment,
we leave $s$ arbitrary). Consider the 
trace of the matrix (\ref{B4})
\begin{eqnarray}
\label{B12}
{\rm tr}\, \Gamma_L &=&{\rm tr}\, \prod_{l=1}^{\vert L\vert} P_{\mu_l k_l}
\end{eqnarray}
and write, say, $P_{\mu_1 k_1}$ in accordance with the formula 
\begin{eqnarray}
\label{B13}
P_{\mu k}&=&\sum^s_{i=1} \phi_{(\mu k,i)}
\otimes\phi^\dagger_{(\mu k,i)}
\end{eqnarray}
where the set of $\phi_{(\mu k,i)}\in V_{\mu k}$, 
$(\phi_{(\mu k,i)},\phi_{(\mu k,j)})_{\bf C} =\delta_{ij}$, 
$i,j=1,\ldots,s$,
is an orthonormal system in the subspace of $V$ described by $P_{\mu k}$.
Then, we can alternatively write for eq.\ (\ref{B12})
\begin{eqnarray}
\label{B14a}
{\rm tr}\, \Gamma_L &=&\tau^{C(L)/2}\ \sum^s_{i=1}\ 
(\phi_{(\mu_1 k_1,i)},\phi^\prime_{(\mu_1 k_1,i)})_{\bf C}\ ,\\
\label{B14b}
\phi^\prime_{(\mu_1 k_1,i)}&=&\tau^{-C(L)/2}\ 
\prod_{l=1}^{C(L)} P_{\mu_l k_l}\ \phi_{(\mu_1 k_1,i)}\ .
\end{eqnarray}
By virtue of the eqs.\ (\ref{CL11}), (\ref{CL12}), the vectors 
$\phi^\prime_{(\mu_1 k_1,i)}\in V_{\mu_1 k_1}$, $i=1,\ldots,s$, 
also represent an orthonormal system in the subspace $V_{\mu_1 k_1}$, i.e.,
$(\phi^\prime_{(\mu_1 k_1,i)},\phi^\prime_{(\mu_1 k_1,j)})_{\bf C} 
=\delta_{ij}$. 
The isocliny of the subspaces described by the projectors
$P_{\mu\pm}$ allows us to write eq.\ (\ref{B14a}) in the following
form. 
\begin{eqnarray}
\label{B15}
{\rm tr}\, \Gamma_L &=&s\ \tau^{C(L)/2}\ 
(\phi_{(\mu_1 k_1,1)},\phi^\prime_{(\mu_1 k_1,1)})_{\bf C}
\end{eqnarray}
To see that each term in the sum on the r.h.s.\ of eq.\ (\ref{B14a})
contributes the same amount, insert for each projector $P_{\mu_l k_l}$
on the r.h.s.\ of eq.\ (\ref{B14b}) the representation (\ref{B13})
and always choose 
$\phi_{(\mu_{l-1} k_{l-1},i)} = \tau^{-1/2} P_{\mu_{l-1} k_{l-1}} 
\phi_{(\mu_l k_l,i)}$.\\

The scalar product
$(\phi_{(\mu_1 k_1,1)},\phi^\prime_{(\mu_1 k_1,1)})_{\bf C}$ 
in eq.\ (\ref{B15}) is nothing 
else than the cosine of the complex angle between the (unit) vectors
$\phi_{(\mu_1 k_1,1)}$ and $\phi^\prime_{(\mu_1 k_1,1)}$ (for a 
review on the subject of angles in complex vector spaces and 
definitions of the angle concepts used here see \cite{scha5}).
First, choose $s=1$. Then, one can write ($\varphi_L\in {\bf R}$,
$-\pi\le\varphi_L\le\pi$).
\begin{eqnarray}
\label{B16}
\phi^\prime_{(\mu_1 k_1,1)}&=&{\rm e}^{\displaystyle i\varphi_L}\ 
\phi_{(\mu_1 k_1,1)}
\end{eqnarray}
and, consequently,
\begin{eqnarray}
\label{B17}
(\phi_{(\mu_1 k_1,1)},\phi^\prime_{(\mu_1 k_1,1)})_{\bf C}
&=&{\rm e}^{\displaystyle i\varphi_L}\ .
\end{eqnarray}
$\varphi_L$ is the pseudo-angle between the vectors $\phi_{(\mu_1 k_1,1)}$ and $\phi^\prime_{(\mu_1 k_1,1)}$. Their Hermitian angle is zero. 
Thus, eq.\ (\ref{B15}) finally reads for $s=1$
\begin{eqnarray}
\label{B18}
{\rm tr}\, \Gamma_L &=&\tau^{C(L)/2}\ {\rm e}^{\displaystyle i\varphi_L}\ .
\end{eqnarray}
Now, choose $s=2$. We apply a real representation of
the Clifford algebra $C(3,0)$ by going over from the complex two-dimensional 
spinor space $V=V_{\bf C}\simeq {\bf C}_2$ for $s=1$ 
to the corresponding four-dimensional real spinor space 
$V=V_{\bf R}\simeq {\bf R}_4$ for $s=2$ which is isometric to the former. 
In accordance with the almost complex structure defined in 
$V_{\bf R}$ by means of eq.\ (\ref{CL13}), a
vector $\Phi\in V_{\bf R}$ is related to a vector 
$\phi=(\phi_1,\phi_2)^T \in V_{\bf C}$ by the formula
\begin{eqnarray}
\label{B19}
\Phi^T&=&
\left({\rm Re}\ \phi_1,{\rm Im}\ \phi_1,
{\rm Re}\ \phi_2,{\rm Im}\ \phi_2\right)\ .
\end{eqnarray}
As the eigenspaces of the gamma matrices in the chosen real
representation of the Clifford algebra $C(3,0)$ are holomorphic
2-planes (cf.\ eq.\ (\ref{B16})), 
their K\"ahler angle is zero and, consequently, the 
(Euclidean) angle between the vectors $\Phi_{(\mu_1 k_1,1)}$ 
and $\Phi^\prime_{(\mu_1 k_1,1)}$ is equal to their pseudo-angle
(cf.\ sect.\ 4 of \cite{scha5}). Thus, for $s=2$ eq.\ (\ref{B15}) reads 
\begin{eqnarray}
\label{B20}
{\rm tr}\, \Gamma_L &=&2\ \tau^{C(L)/2}\ \cos\varphi_L\ .
\end{eqnarray}
From eq.\ (\ref{B11}) we recognize that $\varphi_L =\theta_L/2$.
The eqs.\ (\ref{B18}) and (\ref{B20}) provide us with the 
precise relation between the Dirac traces taken in
the $s=1$ and the $s=2$ representations
of the Clifford algebra $C(3,0)$. Therefore, in the remainder of this
subsection for simplicity we confine our consideration to the case $s=1$.\\

We first study eq.\ (\ref{CL37}). In view of eq.\ 
(\ref{CL35}), we set $k,l = + (1)$. For $s=1$, we can write 
$w_{\mu +}$, $\mu=1,2,3$, as follows (cf.\ eq.\ (\ref{CL36}); 
$\beta_\mu\in {\bf R}$).
\begin{eqnarray}
\label{B21}
w_{\mu +}&=& \sin 2\theta_\mu\ {\rm e}^{\displaystyle i\beta_\mu}
\end{eqnarray}
This leads to these three equations 
($\Delta\beta_{\mu\nu} = \beta_\mu -\beta_\nu$;
we assume here $\theta_\mu \not= 0, \frac{\pi}{2}$, these cases
have to be studied separately).
\begin{eqnarray}
\label{B22a}
\cos\Delta\beta_{12}&=&-\ \cot 2\theta_1\ \cot 2\theta_2\\
\label{B22b}
\cos\Delta\beta_{13}&=&-\ \cot 2\theta_1\ \cot 2\theta_3\\
\label{B22c}
\cos\Delta\beta_{23}&=&-\ \cot 2\theta_2\ \cot 2\theta_3
\end{eqnarray}
The system of eqs.\ (\ref{B22a})-(\ref{B22c}) can be transformed
to read
\begin{eqnarray}
\label{B23a}
\cot^2 2\theta_1&=&-\ 
\frac{\cos\Delta\beta_{12}\ \cos\Delta\beta_{13}}{\cos\Delta\beta_{23}}\ ,\\
\label{B23b}
\cot^2 2\theta_2&=&-\ 
\frac{\cos\Delta\beta_{12}\ \cos\Delta\beta_{23}}{\cos\Delta\beta_{13}}\ ,\\
\label{B23c}
\cot^2 2\theta_3&=&-\ 
\frac{\cos\Delta\beta_{23}\ \cos\Delta\beta_{13}}{\cos\Delta\beta_{12}}\ .
\end{eqnarray}
One can convince oneself by explicit calculation that for any 
(admissible) choice of
$\beta_\mu$, $\mu=1,2,3$, the isocliny angles $\theta_\mu$ given by the eqs.\ 
(\ref{B23a})-(\ref{B23c}) respect the constraint eq.\ (\ref{CL14}) ($p=3$). 
Conversely, any (admissible)
choice of the isocliny angles $\theta_\mu$ determines 
$\beta_\mu$ (up to a $\mu$ independent constant and other obvious 
symmetries of the eqs.\ (\ref{B22a})-(\ref{B22c})). 
Inserting eq.\ (\ref{B21}) into eq.\ (\ref{CL49b})
and taking into account eqs.\ (\ref{B22a})-(\ref{B22c})
we find 
\begin{eqnarray}
\label{B26a}
I_{(\mu k,\nu l)}&=&i\ .
\end{eqnarray}
To arrive at this result, 
\begin{eqnarray}
\label{B26aa}
f_{\mu\nu}&=&{\rm sgn}\left[\sin\left(\beta_\mu -\beta_\nu\right)\right]
\end{eqnarray}
has been chosen. In the following, we will consider two 
different special solutions of the eqs.\ (\ref{B22a})-(\ref{B22c}). 
The first one we will refer to as the symmetric case
while we will call the second one the Pauli case.\\

The symmetric case is approached as follows. For our purpose
of calculating Dirac traces within lattice quantum field theory, it 
seems to be most natural and appropriate to choose $\beta_\mu$,
$\theta_\mu$ in a most symmetric way. Therefore, taking into
account eq.\ (\ref{CL14}) we set 
\begin{eqnarray}
\label{B24}
\theta_1&=&\theta_2\ =\ \theta_3\ =\ \theta^{\{\pm\}}_{\rm sym}\ ,\ \ \ 
\cos 2\theta^{\{\pm\}}_{\rm sym}\ =\ \pm\frac{1}{\sqrt{3}}\ .
\end{eqnarray}
This choice entails (in both cases)
\begin{eqnarray}
\label{B25}
\beta_\mu&=&\beta(\mu)\ =\ \beta_0\ +\ \frac{2\pi}{3}\ \mu
\end{eqnarray}
($\beta_0\in{\bf R}$ is some arbitrary constant). Inserting this
choice into eq.\ (\ref{CL33b}) (select $\theta^{\{+\}}_{\rm sym}$),
we immediately see that these gamma matrices agree 
(up to complex conjugation) with the transformed Pauli matrices 
found in \cite{scha4}, sect.\ I, eq.\ (11), p.\ 3618.
Proceeding further, from eqs.\ (\ref{CL49b})-(\ref{CL49d}), 
(\ref{B24}) we find\footnote{Incidentally, the appearance of 
$\frac{\pi}{12}=\frac{2\pi}{24}$ as value of the 
angles $\alpha^{\{\pm\}}_{(\mu k,\nu l)}$ is eye-catching to some
extent as this angle does not occur very often in mathematics.
One is immediately lead to think of the Dedekind $\eta$ function
where it also occurs. If this signals more than an accidental
coincidence has to remain open for now.}
\begin{eqnarray}
\label{B26b}
\alpha^{\{\pm\}}_{(\mu +,\nu +)}&=& 
f_{\mu\nu}\ \left\{
\begin{array}{*{1}{c}}
1\\ 
7
\end{array}
\right\}\frac{\pi}{12}\ ,\\
\label{B26c}
\alpha^{\{\pm\}}_{(\mu +,\nu -)}&=&\alpha^{\{\pm\}}_{(\mu -,\nu +)}\ =\ 
-\ f_{\mu\nu}\ \frac{\pi}{6}\ ,\\
\label{B26d}
\alpha^{\{\pm\}}_{(\mu -,\nu -)}&=&
f_{\mu\nu}\ \left\{
\begin{array}{*{1}{c}}
7\\ 
1
\end{array}
\right\}\frac{\pi}{12}\ .
\end{eqnarray}
Consequently, we can express $\varphi_L$ as follows.
\begin{eqnarray}
\label{B27}
\varphi_L&=&\frac{\theta_L}{2}\ \equiv\ 
\left(\sum^{C(L)}_{l=1}\ \alpha^{\{\pm\}}_{(\mu_l k_l,\mu_{l+1} k_{l+1})}
\right) {\rm mod}(2\pi)
\end{eqnarray}
In other words, each loop corner occurring in the hopping parameter 
expansion of Wilson fermions in 3 lattice dimensions can be 
attached an angle $\alpha^{\{\pm\}}_{(\mu k,\nu l)}$
whose linear sum determines the value of ${\rm tr}\, \Gamma_L$. The
abelian character of this sum
should be ideally suited for any future study of the 
generalization of eq.\ (\ref{B5}) to 3 lattice dimensions (the same
comment applies to the Pauli case below).\\

We define the Pauli case by choosing 

\parbox{7cm}{
\begin{eqnarray}
\label{B29a}
\gamma_1&=&\sigma_1\ =\ 
\left(
\begin{array}{*{2}{c}}
0&1\\ 
1&0
\end{array}
\right)
\end{eqnarray}
}
\hfill
\parbox{7cm}{
\begin{eqnarray}
\label{B29b}
\gamma_2&=&\sigma_2\ =\ 
\left(
\begin{array}{*{2}{c}}
0&-i\\ 
i&0
\end{array}
\right)
\end{eqnarray}
}

\begin{eqnarray}
\label{B29c}
\gamma_3&=&\sigma_3\ =\ 
\left(
\begin{array}{*{2}{c}}
1&0\\ 
0&-1
\end{array}
\right)
\end{eqnarray}
where $\sigma_k$, $k=1,2,3$, are the standard Pauli matrices.
This choice entails
\begin{eqnarray}
\label{B30a}
\theta_1&=&\frac{\pi}{4}\ ,\ \ \ w_{1+}\ =\ \ \, 1\ ,\ \ \ \beta_1\ =\ 0\ ,\\
\label{B30b}
\theta_2&=&\frac{\pi}{4}\ ,\ \ \ w_{2+}\ =\ -i\ ,\ \ \ 
\beta_2\ =\ -\frac{\pi}{2}\ ,\\
\label{B30c}
\theta_3&=&\ 0\ ,\ \ \ w_{3+}\ =\ 0\ .
\end{eqnarray}
Of course, $\beta_3$ remains undetermined and $U_{3-}$ is not 
immediately given by eq.\ (\ref{CL39}), only after some closer
inspection. We find
\begin{eqnarray}
\label{B31}
U_{3-}&=&
\left(
\begin{array}{*{2}{c}}
0&-\ {\rm e}^{\displaystyle i\beta_3}\\ 
{\rm e}^{\displaystyle -i\beta_3}&0
\end{array}
\right)\ .
\end{eqnarray}
The matrices $\check{t}_{(\mu k,\nu l)}$ can again be found 
without any problem from eqs.\ (\ref{CL49a})-(\ref{CL49d}),
except for $\mu=3$, $k=-(1)$ (or $\nu=3$, $l=-(1)$). To proceed
in the latter case on the basis of the eqs.\ (\ref{CL49a})-(\ref{CL49d})
involves a subtle limiting procedure for the isocliny angles
$\theta_{\mu k}$. The limit for $\theta_1$, $\theta_2$, $\theta_3$
towards $\frac{\pi}{4}$ and $0$ is ruled by the constraint eq.\ (\ref{CL14})
($p=3$) and the value of $\beta_3$ has to be adjusted in accordance
with the chosen approach towards this limit 
on the basis of the eqs.\ (\ref{B22b}), (\ref{B22c}). 
For $\mu=3$, $k=-(1)$ (or $\nu=3$, $l=-(1)$), a direct calculation
is simpler and yields ($\nu = 1,2$)
\begin{eqnarray}
\label{B32}
\check{t}_{(3 -,\nu l)}&=&\check{t}^\dagger_{(\nu l,3 -)}\ =\
\sqrt{2}\ \left[ U_{3-} U^\dagger_{\nu l}
\right]_{11}\ =\ -\ l\ {\rm e}^{\displaystyle i(\beta_3 -\beta_\nu)}\ .
\end{eqnarray}
Finally, we find (in the present case 
we have chosen $f_{12} = f_{13} = f_{23} = 1$)
\begin{eqnarray}
%\label{B33a}
\label{B33b}
\alpha_{(1 k,2 l)}&=&-\ \alpha_{(2 l,1 k)}\ =\ k\ l\ 
f_{12}\ 
\frac{\pi}{4}\ ,\\
\label{B33c}
\alpha_{(\nu l,3+)}&=&\alpha_{(3 +,\nu l)}\ =\ 0\ ,\\
\label{B33d}
\alpha_{(\nu +,3-)}&=&-\ \alpha_{(3-,\nu +)}\ =\ 
f_{\nu 3}\ (\beta_\nu -\beta_3 +\pi)\ ,\\
\label{B33e}
\alpha_{(\nu -,3-)}&=&-\ \alpha_{(3-,\nu -)}\ =\
f_{\nu 3}\ (\beta_\nu -\beta_3)\ .
\end{eqnarray}
Again, $\varphi_L$ can be expressed the same way as in eq.\
(\ref{B27}).
\begin{eqnarray}
\label{B34}
\varphi_L&=&\frac{\theta_L}{2}\ \equiv\ 
\left(\sum^{C(L)}_{l=1}\ \alpha_{(\mu_l k_l,\mu_{l+1} k_{l+1})}
\right) {\rm mod}(2\pi)
\end{eqnarray}
For simplicity, we can set $\beta_3=0$. Then, eqs.\ 
(\ref{B33b})-(\ref{B34}) immediately tell us that in 3 lattice 
dimension $\theta_L$ must be a multiple of $\frac{\pi}{2}$
as argued in \cite{stam} by other means. Also, a another rule
given in \cite{stam} can independently be rederived (also cf.\ 
the comments made in the present paper below of eq.\ (\ref{B11})).
Replace in $\Gamma_L$ (eq.\ (\ref{B4})) all projectors $P_{2\pm}$
by projectors $P_{1\pm}$ (it does not matter for the present argument
if the procedure makes any geometrical sense). Then, for the modified trace 
${\rm tr}\, \Gamma_{L^\prime}$ the corner number $C(L^\prime)$ 
is always even and all corner angles
are given by eqs.\ (\ref{B33c})-(\ref{B33e}). Consequently, 
$\theta_{L^\prime}$ is an even multiple of $\frac{\pi}{2}$. 
Therefore, the number of corners with $\mu_l =1$, $\mu_{l+1} =2$, or
$\mu_l =2$, $\mu_{l+1} =1$, determines if $C(L)$ is even or odd. From
this fact and eq.\ (\ref{B33b}) we can immediately conclude that
$\theta_L$ can only be an odd multiple of $\frac{\pi}{2}$ if the 
corner number $C(L)$ is also odd.\\

Finally, let us shortly comment on the Dirac traces taken in 
the $s=2$ representation of the Clifford algebra $C(3,0)$. 
According to eq.\ (\ref{CL13}), eq.\ (\ref{B26a}) then reads
\begin{eqnarray}
\label{B28}
I_{(\mu k,\nu l)}&=&
\left(
\begin{array}{*{2}{c}}
0&-1\\ 
1&0
\end{array}
\right)\ .
\end{eqnarray}
Consequently, within the present approach
the matrices (\ref{CL49a}) related to the 
loop corners are $SO(2)$ matrices ($\sigma_2$
denotes the second Pauli matrix)
\begin{eqnarray}
\label{B35}
\check{t}_{(\mu k,\nu l)}
&=&
\left(
\begin{array}{*{2}{c}}
\cos\alpha_{(\mu k,\nu l)}&-\sin\alpha_{(\mu k,\nu l)}\\ 
\sin\alpha_{(\mu k,\nu l)}&\cos\alpha_{(\mu k,\nu l)}
\end{array}
\right)
\ =\ {\rm e}^{\displaystyle - i \alpha_{(\mu k,\nu l)} \sigma_2}
\ .
\end{eqnarray}
For $s=2$, the gamma matrices given by eq.\ (\ref{CL33b}) 
agree in the symmetric case 
(up to some elementary transformation, i.e., an inversion) with those
given in \cite{scha4}, sect.\ I, eq.\ (7), p.\ 3617.\\

\subsection{Dirac traces in 4 lattice dimensions}

Unfortunately, the strategy applied in the previous subsection
for 3 lattice dimensions cannot be extended to 4 dimensions.
The reason for this consists in the fact that there is no
$s=1$ matrix representation of the Clifford algebra $C(4,0)$. On a technical
level, in our approach this fact raises its head 
if one would attempt to solve the analogue
of eqs.\ (\ref{B22a})-(\ref{B22c}) for 4 dimensions.
This set of equations would then consist of 6
equations which are no longer simultaneously solvable.
Consequently, in 4 lattice dimensions we have to work right 
with a $s=2$ representation of the gamma matrices.  Now, what is the best 
strategy in 4 lattice dimensions? We do not have any final answer
on this question but it seems not unreasonable to assume for the moment
that the approach should be based on the 
results obtained in the previous subsection for 3 dimensions.\\

First, we have to study eq.\ (\ref{CL37}) again. As 
in the chosen approach we have in accordance with eq.\ (\ref{B21}) 
\begin{eqnarray}
\label{B36}
w_{\mu +}&=&\sin 2\theta_\mu\
\left(
\begin{array}{*{2}{c}}
\cos\beta_\mu&-\sin\beta_\mu\\ 
\sin\beta_\mu&\cos\beta_\mu
\end{array}
\right)\ ,\ \ \  \mu = 1,2,3,
\end{eqnarray}
(in particular, this equation applies in both the symmetric and 
the Pauli cases but not only then)
we immediately find from eq.\ (\ref{CL37}) (clearly, in view of 
eq.\ (\ref{CL14}) $\theta_4 =\frac{\pi}{4}$)
\begin{eqnarray}
\label{B37}
w_{4 +}&=&i\
\left(
\begin{array}{*{2}{c}}
\cos\beta_4&\sin\beta_4\\ 
\sin\beta_4&-\cos\beta_4
\end{array}
\right)\ .
\end{eqnarray}
Consequently, the matrix $-i w_{4 +}$ belongs to the group
$O(2)$. $\beta_4 \in {\bf R}$ can freely be chosen.
From eqs.\ (\ref{CL49a})-(\ref{CL49d}) we find
($\sigma_2$, $\sigma_3$ are the standard Pauli matrices)
\begin{eqnarray}
\label{B38}
\check{t}_{(\mu k,4 l)}
&=&\check{t}^\dagger_{(4 l,\mu k)}\nonumber\\
&=&\cos\theta_{\mu k} - i \sin\theta_{\mu k}\ 
\left(
\begin{array}{*{2}{c}}
\cos(\beta_\mu +\beta_4)&\sin(\beta_\mu +\beta_4)\\ 
\sin(\beta_\mu +\beta_4)&-\cos(\beta_\mu +\beta_4)
\end{array}
\right)\\
&=&
\left(
\begin{array}{*{2}{c}}
\cos\frac{\beta_\mu +\beta_4}{2}&-\sin\frac{\beta_\mu +\beta_4}{2}\\ 
\sin\frac{\beta_\mu +\beta_4}{2}&\cos\frac{\beta_\mu +\beta_4}{2}
\end{array}
\right)\ 
\left(
\begin{array}{*{2}{c}}
{\rm e}^{\displaystyle -i \theta_{\mu k}}&0\\ 
0&{\rm e}^{\displaystyle i \theta_{\mu k}}
\end{array}
\right)\nonumber\\
&&\ \ \ \times\ 
\left(
\begin{array}{*{2}{c}}
\cos\frac{\beta_\mu +\beta_4}{2}&\sin\frac{\beta_\mu +\beta_4}{2}\\ 
-\sin\frac{\beta_\mu +\beta_4}{2}&\cos\frac{\beta_\mu +\beta_4}{2}
\end{array}
\right)\nonumber\\
&=&
\exp\left[- i\ \frac{\beta_\mu +\beta_4}{2}\ \sigma_2\right]\ 
\exp\left[- i \theta_{\mu k} \sigma_3\right]\ 
\exp\left[ i\ \frac{\beta_\mu +\beta_4}{2}\ \sigma_2\right]\ .
\end{eqnarray}
The last line can be understood as a representation of the 
$SU(2)$ matrix $\check{t}_{(\mu k,4 l)}$ in terms of Euler angles
(cf., e.g., \cite{bied}, chap.\ 2, sect.\ 7, eq.\ (2.40), p.\ 24).
We can then write
\begin{eqnarray}
\label{B39}
\check{t}_{(\mu k,4 l)} \check{t}_{(4 l,\nu m)} 
&=&
\exp\left[- i\ \frac{\beta_\mu +\beta_4}{2}\ \sigma_2\right]\ 
\exp\left[- i \theta_{\mu k} \sigma_3\right]\ 
\exp\left[ i\ \frac{\beta_\mu -\beta_\nu}{2}\ \sigma_2\right]\nonumber\\
&&\times\ \exp\left[ i \theta_{\nu m} \sigma_3\right]\ 
\exp\left[ i\ \frac{\beta_\nu +\beta_4}{2}\ \sigma_2\right]\ .
\end{eqnarray}
From the above equations one recognizes that in 4 lattice 
dimensions the non-abelian character of the product appearing 
on the r.h.s.\ of eq.\ (\ref{B9}) is closely related to the values
of the isocliny angles $\theta_{\mu k}$. Introducing the 
notation (we assume without restricting generality 
$\theta_{\mu +}\le\frac{\pi}{4}$)
\begin{eqnarray}
\label{B40}
\theta_{\mu k}&=&\frac{\pi}{4}\ -\ k\ \Delta\theta_\mu
\end{eqnarray}
we can express the product of the three middle matrices 
on the r.h.s.\ of eq.\ (\ref{B39}) which encodes the non-abelian 
structure as follows (here, $\omega_{\mu\nu} = (\beta_\mu-\beta_\nu)/2$).
\begin{eqnarray}
\label{B41}
&&\hspace{-4.5cm}\exp\left[- i \theta_{\mu k} \sigma_3\right]\
\exp\left[ i\ \omega_{\mu\nu}\ \sigma_2\right]\
\exp\left[ i \theta_{\nu m} \sigma_3\right]\nonumber\\
&=&\cos\left(k \Delta\theta_\mu - m \Delta\theta_\nu\right)\
\cos\omega_{\mu\nu}\nonumber\\
&&-\ i \sigma_1\ \sin\omega_{\mu\nu}\ 
\cos\left(k \Delta\theta_\mu + m \Delta\theta_\nu\right)\nonumber\\ 
&&+\ i \sigma_2\ \sin\omega_{\mu\nu}\ 
\sin\left(k \Delta\theta_\mu + m \Delta\theta_\nu\right)\nonumber\\
&&+\ i \sigma_3\ \cos\omega_{\mu\nu}\ 
\sin\left(k \Delta\theta_\mu - m \Delta\theta_\nu\right)
\end{eqnarray}
From this expression, one easily recognizes that 
it would be advantageous if $\Delta\theta_\mu = 0$
(i.e., $\theta_{\mu k} = \frac{\pi}{4}$) applied 
for as many as possible values of $\mu$
in order to facilitate and to simplify
the further study of Dirac traces in 4 lattice dimensions. 
It is clear, that in view of eq.\ (\ref{CL14}) 
$\Delta\theta_1 = \Delta\theta_2 =\Delta\theta_3 =0$
is not an admissible choice.
There are two alternatives to this best, but inadmissible choice.
Either one applies the condition
$\Delta\theta_1 = \Delta\theta_2 =\Delta\theta_3$ ($\not= 0$)
which leads to the symmetric case or, one chooses 
$\Delta\theta_1 = \Delta\theta_2 =0$ which leads to the Pauli case
(also, considering $\omega_{\mu\nu}$ does not lead to
any better choices as the values of $\omega_{\mu\nu}$ are closely
related to the isocliny angles $\theta_\mu$ according to eqs.\
(\ref{B22a})-(\ref{B22a})). Studying in both cases the explicit matrix 
expressions (\ref{B41}), unfortunately neither the symmetric 
nor the Pauli case seem to exhibit any particular advantage
for their application in the further study of Dirac traces in 
4 lattice dimensions. Still, alternatively one might try to further
uphold the idea that possibly best further progress could be made
if all eight isocliny angles
$\theta_{\mu \pm}$ one has to deal with in 4 dimensions 
would have the same value. This can only be achieved
by extending the consideration in the Pauli case
to the Clifford algebra $C(5,0)$
(which is the maximal Clifford algebra with a $s=2$ representation).
Then, eight isocliny angles $\theta_{\mu \pm}$ can assume the 
value $\frac{\pi}{4}$.  From eq.\ (\ref{B37}) we immediately find
(again, eq.\ (\ref{CL14}) tells us that $\theta_5 =\frac{\pi}{4}$)
\begin{eqnarray}
\label{B42}
w_{5 +}&=&i\
\left(
\begin{array}{*{2}{c}}
\cos\beta_5&\sin\beta_5\\ 
\sin\beta_5&-\cos\beta_5
\end{array}
\right)
\end{eqnarray}
and eq.\ (\ref{CL37}) (which agrees in the Pauli case with eq.\ 
(\ref{CL45})) yields 
\begin{eqnarray}
\label{B43}
\beta_4 -\beta_5&\equiv&\frac{\pi}{2}\ {\rm mod}(\pi)\ .
\end{eqnarray}
To be specific, we choose
\begin{eqnarray}
\label{B44a}
\beta_4&=&0\ ,\\
\label{B44b}
\beta_5&=&\frac{\pi}{2}\ .
\end{eqnarray}
Our choices result in the following $s=2$ representation of the 
Clifford algebra $C(4,0)$ 
($\theta_{1\pm}=\theta_{2\pm}=\theta_{4\pm}=\theta_{5\pm}=\frac{\pi}{4}$).
We only display the matrices $w_{\mu +}$, for the corresponding 
gamma matrices see eq.\ (\ref{CL33b}) ($\sigma_k$ are the standard Pauli
matrices).
\begin{eqnarray}
\label{B45a}
w_{1 +}&=&{\bf 1}_2\\
\label{B45b}
w_{2 +}&=&
\left(
\begin{array}{*{2}{c}}
0&1\\ 
-1&0
\end{array}
\right)\ =\ i \sigma_2\\
\label{B45c}
w_{4 +}&=&i\
\left(
\begin{array}{*{2}{c}}
1&0\\ 
0&-1
\end{array}
\right)\ =\ i \sigma_3\\
\label{B45d}
w_{5 +}&=&i\
\left(
\begin{array}{*{2}{c}}
0&1\\ 
1&0
\end{array}
\right)\ =\ i \sigma_1
\end{eqnarray}
Obviously, this consideration has brought us back to the 
chiral representation of the gamma matrices which had already 
been used in \cite{stam} (cf.\ eqs.\ (\ref{B7a})-(\ref{B7d})).
At the level of the present study, it seems difficult to 
make any final judgement which representation of Dirac traces
for Wilson fermions in 4 lattice dimensions will facilitate
their further study in the most effective way. Consequently,
the current subsection could be best understood as an
explorative one.\\

\section{Final comments}

The present study has revealed that 
there is a considerable difference between Dirac traces for Wilson fermions
in 3 and 4 lattice dimensions. While in 3 dimensions each 
loop corner occurring in the hopping parameter expansion for 
Wilson fermions an angle can be associated with whose linear
sum (over all loop corners) determines the value of the 
corresponding trace such a simple procedure cannot be 
introduced in 4 lattice dimensions. This follows in a quite 
straightforward way from the study of Clifford algebra
representations performed in sect.\ 2. On the basis of the 
parametrization of gamma matrices in terms of isocliny angles,
we have discussed some of the possible representations of 
Dirac traces in 4 dimensions. The formalism developed in the 
present paper allows in principle to systematically map out
the space of such representations and, therefore, should be useful in 
any future study of the subject. It remains to be hoped that
the progress achieved in the present paper will not only allow
to gain further future understanding of Dirac traces for Wilson
fermions in 3 lattice dimensions but also to successfully
attack the problem in 4 dimensions.\\

At the end of this study, we would like to mention some 
further related articles.
Wilson fermions can be understood as a statistical system with 
matrix-valued ($SU(2)$) vertex weights (and a bending 
rigidity $\eta = 1/\sqrt{2}$; cf.\ eqs.\ (\ref{B3}), (\ref{B9})). 
Models of the same type have been considered in the past, e.g., 
in the approximate study of the three-dimensional Ising
model \cite{shih1}-\cite{cont} (in these papers the 
authors apply $SU(2)$ vertex
weights and a bending rigidity $\eta = 1$). A variable bending 
rigidity has been studied in a four-dimensional lattice fermion model with 
(Wilson fermion) $SU(2)$ matrix vertex weights in \cite{brow1,brow2}
(`link fermions').
Also, recently a three-dimensional loop model with $SU(2)$ vertex
weights and a variable bending rigidity $\eta$ has been considered
\cite{scho} (however, the authors of this work pay special attention 
to the bending rigidity $\eta = 1/\sqrt{2}$ which is the case
most closely related to Wilson fermions; in this context it should
be pointed out that only further research can show how eq.\ 
(\ref{B6}) is properly generalized to 3 lattice dimensions).
Finally, we want to mention that other three-dimensional loop models with
loop shape dependent weights have also been considered
\cite{itoi}-\cite{higu}.\\

\end{document}